\documentclass[letterpaper,preprintnumbers,english,floats,floatfix,amssymb,prd,superscriptaddress,twocolumn,nofootinbib]{revtex4}
\usepackage{amssymb,amsmath}
\usepackage{epsfig,hyperref}
\usepackage[svgnames]{xcolor}
\usepackage{pgf,tikz}
\usepackage{epstopdf}
\usepackage[british]{babel}
\usepackage{enumerate}
\usepackage{enumitem}
\usepackage{makecell}
\usepackage{booktabs}

\makeatletter
\DeclareRobustCommand*{\bfseries}{%
  \not@math@alphabet\bfseries\mathbf
  \fontseries\bfdefault\selectfont
  \boldmath
}
\makeatother

\def\be{\begin{equation}}
\def\ee{\end{equation}}
\def\beq{\begin{eqnarray}}
\def\eeq{\end{eqnarray}}

\makeatletter

\newcommand{\arXiv}[2][]{\href{http://arxiv.org/abs/#2}{\texttt{arXiv:#2\@ifempty{#1}{}{ [#1]}}}}
\makeatother

\begin{document}
\title{Strength of the naked singularity in critical collapse}

\author{Jun-Qi Guo}%
\email{sps{\_}guojq@ujn.edu.cn}
\affiliation{School of Physics and Technology, University of Jinan, Jinan 250022, Shandong, China}

\author{Lin Zhang}%
\email{linzhang2013@pku.edu.cn}
\affiliation{College of Mathematics and Statistics, Chongqing University, Chongqing 401331, China}

\author{Yuewen Chen}%
\email{yuewen{\_}chern@amss.ac.cn}
\affiliation{
 Institute of Applied Mathematics,
 Morningside Center of Mathematics and LESC,
 Institute of Computational Mathematics,
 Academy of Mathematics and System Science, Chinese Academy of Sciences, University of Chinese Academy of Sciences,
Beijing 100190, China.}

\author{Pankaj S. Joshi}%
\email{psjcosmos@gmail.com}
\affiliation{International Center for Cosmology, Charusat University, Anand 388421, Gujarat, India}

\author{Hongsheng Zhang}%
\email{sps{\_}zhanghs@ujn.edu.cn}
\affiliation{School of Physics and Technology, University of Jinan, Jinan 250022, Shandong, China}

\date{\today}

\begin{abstract}
The critical collapse of a scalar field is a threshold solution of black hole formation, in which a naked singularity arises. We study here the curvature strength of this singularity using a numerical ansatz. The behavior of the Jacobi volume forms is examined along a non-spacelike geodesic in the limit of approach to the singularity. These are seen to be vanishing, thus showing that all physical objects will be crushed to zero size near the singularity. Consequently, although the critical collapse is considered to be a fine-tuning problem, the naked singularity forming is gravitationally strong. This implies that the spacetime cannot be extended beyond the singularity, thus making the singularity genuine and physically interesting. These results imply that the nature of critical collapse may need to be examined and explored further.
\end{abstract}
\maketitle

\section{Introduction}
It is widely believed that the vicinities of spacetime singularities are an ideal place to decipher quantum gravity. However, the Weak Cosmic Censorship Conjecture, a long debated issue in general relativity, states that any spacetime singularity should be necessarily hidden inside an event horizon~\cite{Penrose_1969}. If this conjecture is valid, it would be challenging to extract information on quantum gravity from the neighborhoods of singularities directly. On the other hand, if realistic singularities, such as those forming in gravitational collapse of physically reasonable matter fields, were not always covered by an event horizon or gravity, the task of receiving quantum gravity signals from ultra-strong gravity regions in Universe would become easier.

Since the proposal of the Weak Cosmic Censorship Conjecture, several counterexamples to this conjecture have been found, including collapse of dust~\cite{Eardley_1979,Christodoulou_1984,Joshi_1993}, perfect fluids~\cite{Ori_1987,Ori_1990,Joshi_1992} and scalar field~\cite{Christodoulou_1994,Brady:1994aq,Zhang:2015rsa}. The instability of naked singularities in scalar field collapse was analyzed in Refs.~\cite{Christodoulou_1999,Liu_2018}. Motivated by the question posed by Christodoulou, namely whether the smallest black hole forms in scalar field collapse would have finite or infinitesimal mass, Choptuik simulated gravitational collapse of a scalar field~\cite{Choptuik:1992jv,Choptuik:1993}. When the scalar field is weak, the field will implode and then disperse, while a black hole will form when the scalar field is strong enough. By fine-tuning the strength of the scalar field, a critical solution was obtained. The main results of critical collapse analysis include discrete self-similarity, universality and mass-scaling law. The global structure of critical collapse was investigated in Refs.~\cite{Gundlach:1995kd,Gundlach:1996eg,MartinGarcia:2003gz}, and a real analytic solution to critical collapse was  proved to exist in Ref.~\cite{Reiterer:2012mz}. In critical collapse, the scalar field carries away its energy and an infinitesimal naked singularity is left behind~\cite{Scheel:2014hfa}, which is a counterexample to the original censorship conjecture. Due to this discovery, the cosmic censorship was restated such that it only applies to collapse with generic initial data. Consequently, the naked singularity produced in critical collapse was suggested to become less worrisome. For reviews on critical collapse, see Refs.~\cite{Choptuik:1997mq,Gundlach:2007gc,Choptuik:2015mma}.

The gravitational strength is an important characteristic in describing the nature of spacetime singularities. Tipler defined a singularity to be gravitationally strong if an object approaching the singularity is crushed to a zero volume~\cite{Tipler_1977}. In mathematical language, a singularity is called gravitationally strong if the volume (area) element, defined by three (two) independent vorticity-free Jacobi fields via the exterior product along a timelike (null) geodesic from or toward the singularity goes to zero as the singularity is approached. The spacetime cannot be extended beyond a strong singularity, while the extension may be implemented for a weak one. The strength of some singularities has been investigated, including the shell-crossing and shell-focusing singularities in dust collapse~\cite{Newman_1986,Nolan:1999tw,Waugh_1988,Joshi_1993}, naked singularities in perfect-fluid collapse~\cite{Lake_1988,Waugh_1989,Ori_1990,Joshi_1992}, singularities in the Schwarzschild, Friedmann-Robertson-Walker and Kasner solutions and in black hole formation by scalar collapse~\cite{Burko:1997xa}, central singularity in continuous self-similar critical collapse of a scalar field~\cite{Nolan:1999tw,Roberts}, and mass-inflation singularities near the Cauchy horizons of Reissner-Nordstr\"{o}m and Kerr black holes~\cite{Ori_1991,Ori_1992}. For a review on the strength of singularities, see Ref.~\cite{Krolak:1999wk}.

Considering the basic role that the critical collapse might play in general relativity, in this Letter, we analyze the strength of the naked singularity in this model. We find that, although the critical collapse may not occur naturally or could be termed as fine-tuned, the naked singularity forming in this model is gravitationally strong. In this sense, the nature of critical collapse may need to be re-examined and explored further for its possible physical implications. Throughout the Letter, we set $G=c=1$.

\section{Methodology}
We simulate critical collapse of a massless scalar field $\phi$ in spherical symmetry and investigate the strength of the naked singularity in the ansatz,
\be \text{d}s^{2} = e^{-2\sigma(x,t)}(-\text{d}t^2+\text{d}x^2)+{r^{2}(x,t)}\text{d}\Omega^2.\ee
The equations of motion are the following,
\be -r_{,tt}+r_{,xx} - \frac{2m}{r^2}e^{-2\sigma}=0,\label{equation_r}\ee
\be -\sigma_{,tt}+\sigma_{,xx} - \frac{2m}{r^3}e^{-2\sigma}+4\pi(\phi_{,t}^2-\phi_{,x}^2)=0,\label{equation_sigma}\ee
\be -\phi_{,tt}+\phi_{,xx}+\frac{2}{r}(-r_{,t}\phi_{,t}+r_{,x}\phi_{,x})=0,\label{equation_phi}\ee
where the $(_{,t})$ and $(_{,x})$ denote derivatives with respect to the coordinates $t$ and $x$, respectively. $m$ is the Misner-Sharp mass. Details on numerics are depicted in Ref.~\cite{Guo_2018}. The simulation is implemented with second-order finite difference method. The initial profile for the scalar field is set as $\phi(x,0)=a\cdot\exp\left(-{x^2}/{\delta}\right)$. The parameter $a$ is precisely tuned to obtain the numerical solution to critical collapse. Compared to Ref.~\cite{Guo_2018}, in this work we use $m_{,x}$, instead of $m_{,t}$, to obtain $m$, and mesh refinement technique is applied~\cite{Garfinkle:1994jb}.

We explore the curvature strength of the singularity by the aid of one test particle's trajectory, which is a radial timelike geodesic approaching the singularity. The geodesic equations are
\be \ddot{x}-\sigma_{,x}(\dot{t}^2+\dot{x}^2)-2\sigma_{,t}\dot{x}\dot{t}=0,\label{geodesic_eq_x}\ee
\be \ddot{t}-\sigma_{,t}(\dot{t}^2+\dot{x}^2)-2\sigma_{,x}\dot{x}\dot{t}=0,\label{geodesic_eq_t}\ee
\be e^{-2\sigma}(\dot{t}^2-\dot{x}^2)=1,\label{geodesic_eq_constraint}\ee
where the dot $(\dot{})$ denotes derivative with respect to the proper time $\tau$ for the test particle along its geodesic path, and $\tau$ is set to be zero at the center and positive for $r>0$. We solve the geodesic equations with the initial data being fine tuned such that the geodesic terminates at the singularity.

Let $|p|$, $|q|$ and $|b|$ denote the norms of the Jacobi fields, in $\theta$, $\varphi$ and radial directions, respectively. Here $p$ and $q$ satisfy the same evolution equation. From the geodesic deviation equation, the evolution equations for $p$ and $b$ can be obtained~\cite{Nolan:1999tw}
\be r\ddot{p}+2\dot{r}\dot{p}=0,\label{Jacobi_p}\ee
\be \ddot{b}+e^{2\sigma}(\sigma_{,tt}-\sigma_{,xx})b=0.\label{Jacobi_b}\ee
For simplicity, we set $p=q$. Then the norm of the volume element expanded by the Jacobi fields is~\cite{Nolan:1999tw}
\be ||V||=|b|p^{2}r^2.\label{condition_volume}\ee

\begin{figure}[t!]
  \begin{tabular}{ccc}
  \epsfig{file=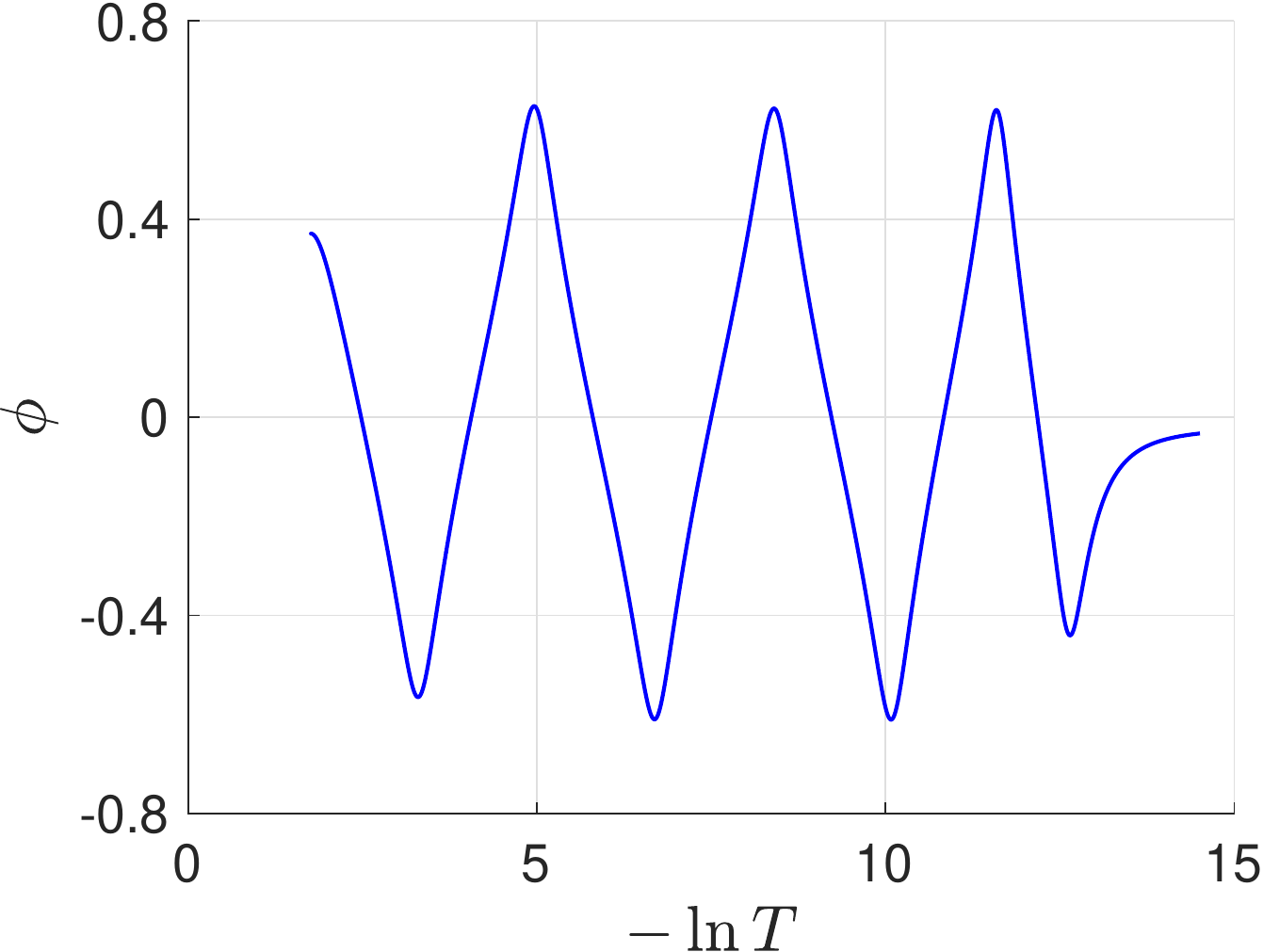, width=7cm}
  \end{tabular}
  \caption{$\phi$ vs. $-\ln{T}$ at the center. $T$ is the proper time for a central observer.}
  \label{fig:phi_vs_tau}
\end{figure}

Based on the condition by Clarke and Krolak~\cite{Clarke_1985}, in non-vacuum collapse, a singularity is strong in the sense of Tipler if along the non-spacelike geodesic, the integral
\be \displaystyle{\int_{c}^0\int_{c}^{\tau_1}R_{\alpha\beta}T^{\alpha}T^{\beta}}(\tau)\text{d}
\tau\text{d}\tau_1\label{condition_strong}\ee
diverges, where $R_{\alpha\beta}$ is the Ricci tensor, $c$ is any fixed positive constant, and $T^{\alpha}(\equiv dx^{\alpha}/d\tau)$ is the tangent vector to the geodesic. When the matter is a scalar field, using $R_{\alpha\beta}=8{\pi}\phi_{,\alpha}\phi_{,\beta}$ and setting $\lambda\equiv-\ln\tau$, we have $R_{\alpha\beta}T^{\alpha}T^{\beta}=8\pi(\phi_{,\tau})^2$ and
\begin{align}
\begin{split}
&\int_{c}^{0}\int_{c}^{\tau_1}(\phi_{,\tau})^2\text{d}\tau\text{d}\tau_1\\
=&\int_{-\ln c}^{+\infty}\left(\int_{-\ln c}^{\lambda_1}e^\lambda\cdot(\phi_{,\lambda})^2\text{d}\lambda\right)\cdot e^{-\lambda_1}\text{d}\lambda_1\\
=&\int_{-\ln c}^{+\infty}(\phi_{,\lambda})^2\text{d}\lambda-\lim\limits_{\lambda_1\to+\infty}
\frac{\int_{-\ln c}^{\lambda_1}e^{\lambda}\cdot(\phi_{,\lambda})^2\text{d}\lambda}{e^{\lambda_1}}.
\end{split}
\nonumber
\end{align}
If $|\phi_{,\lambda}|$ is bounded by some uniformity constant, then the second term in the last line is also bounded by some finite constant. Hence, the divergence of  \eqref{condition_strong} is equivalent to that the integral
\be \int_{-\ln c}^{+\infty}(\phi_{,\lambda})^2\text{d}\lambda \label{condition_strong2}\ee
diverges.

In this work, we first obtained the values for $\sigma_{,x}$, $\sigma_{,t}$ and $r$ etc in Eqs.~(\ref{geodesic_eq_x})-(\ref{Jacobi_b}) and $\phi$ on the grid $(x_{i},t_{i})$ in the simulation. Then, interpolating their values on the geodesic, we solved Eqs.~(\ref{geodesic_eq_x})-(\ref{Jacobi_b}) with the 4th-order Runge-Kutta method, and computed the quantity $||V||$ in (\ref{condition_volume}) and the integral in (\ref{condition_strong2}).
\\

\begin{figure}[t!]
  \begin{tabular}{ccc}
  \epsfig{file=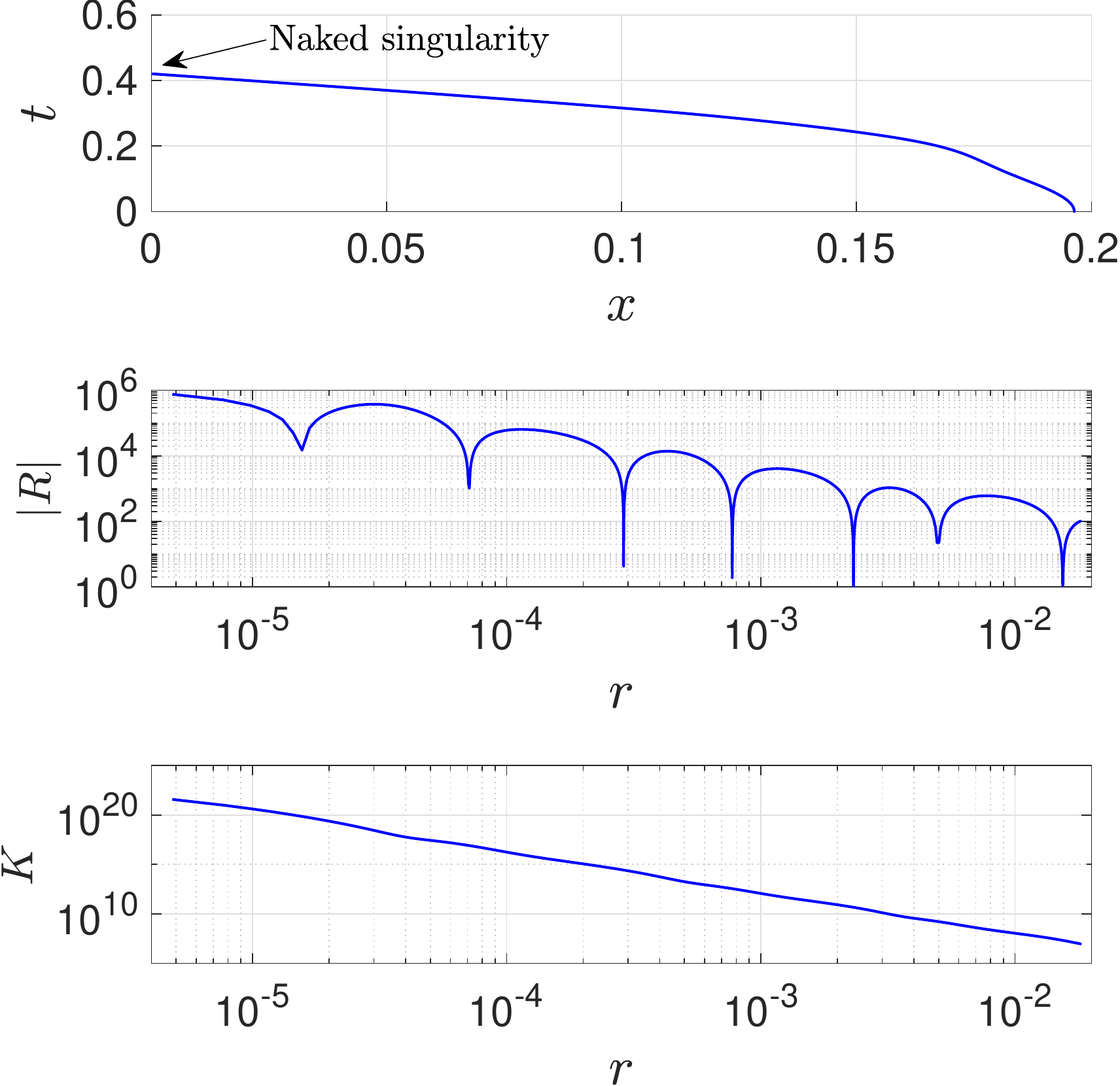, width=7cm}
  \end{tabular}
\caption{Upper panel: one timelike geodesic terminating at the naked singularity. Middle and lower panels: the Ricci scalar $R$ and Kretschmann scalar $K$ along the geodesic.}
  \label{fig:singularity}
\end{figure}

\begin{figure}[t!]
  \begin{tabular}{ccc}
  \epsfig{file=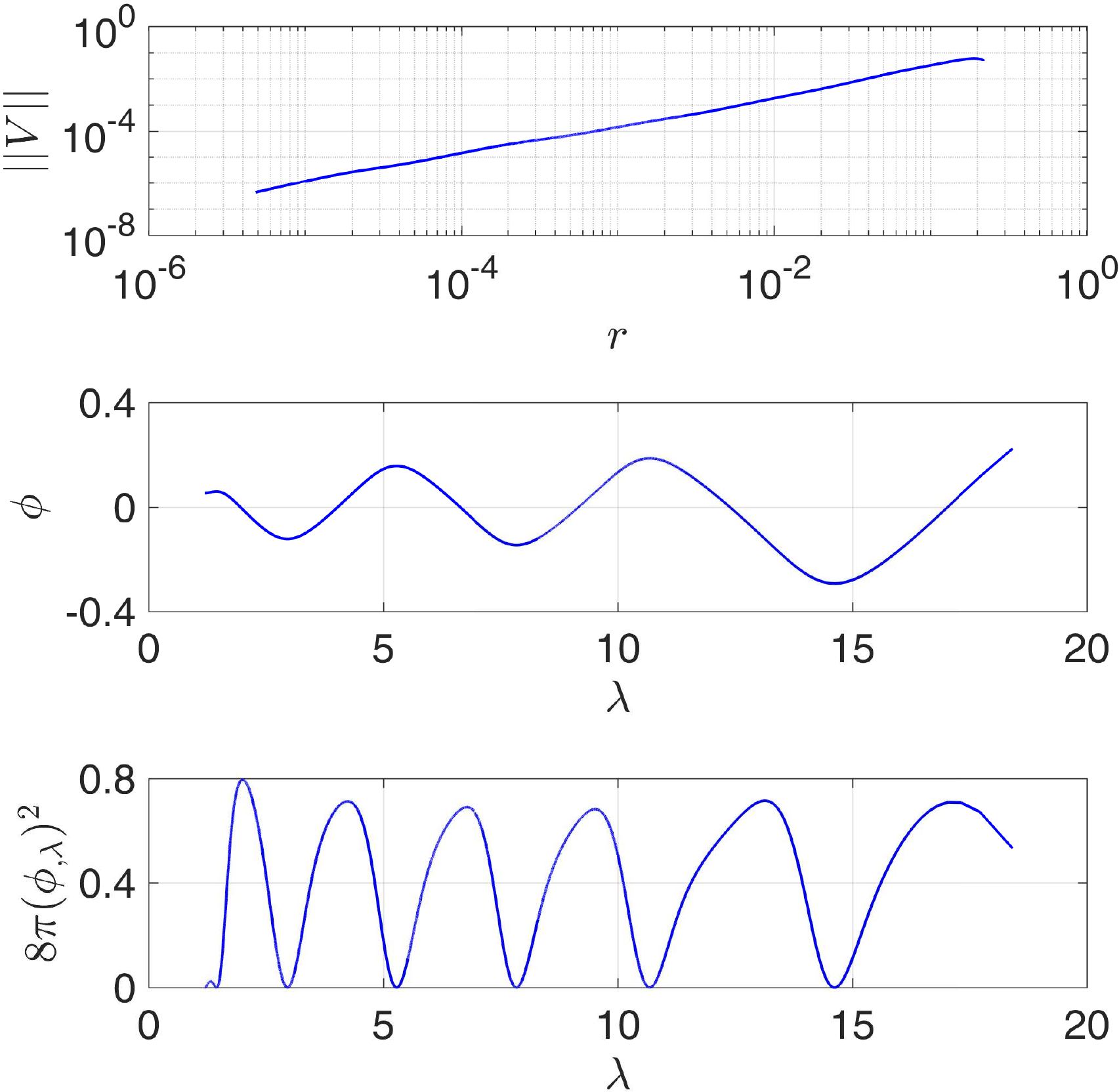, width=7cm}
  \end{tabular}
  \caption{Results on the strength of the singularity. Upper panel: the norm of the volume element $||V||(=|b|p^{2}r^2)$ along the geodesic approaching the singularity. Middle and lower panels: $\phi$ and $8\pi(\phi_{,\lambda})^2[=\tau^{2}R_{\alpha\beta}T^{\alpha}T^{\beta}]$ vs. $\lambda(\equiv-\ln\tau)$ along the geodesic.}
  \label{fig:stength}
\end{figure}

\section{Results and discussions}
We first simulated critical collapse and obtained the numerical values for the spacetime and scalar field. The numerical results for $\phi$ versus $\ln{T}$ at the center are shown in Fig.~\ref{fig:phi_vs_tau}. $T$ is the proper time for a central observer, $T\equiv\int_{0}^{t}e^{-\sigma(0,\tilde{t})}d\tilde{t}$, where $t$ is set to be zero (positive) as (before) the singularity is approached. The results are similar to those in Refs.~\cite{Hamade:1995ce,Choptuik:2003ac,Baumgarte:2018}, where the double-null coordinates, polar-slicing coordinates and Baumgarte-Shapiro-Shibata-Nakamura (BSSN) formalism were used, respectively. Solving Eqs.~(\ref{geodesic_eq_x}) and (\ref{geodesic_eq_t}) with proper initial data for the test particle, we obtained one timelike geodesic terminating at the naked singularity. This is verified by the divergence of the Ricci scalar $R$ and Kretschmann scalar $K$ along the geodesic. See Fig.~\ref{fig:singularity}.

Via integrations of Eqs.~(\ref{Jacobi_p}) and (\ref{Jacobi_b}), the norm of the volume element $||V||$ by the Jacobi fields was obtained. As shown in the upper panel of Fig.~\ref{fig:stength}, $||V||$ vanishes as the center is approached. Therefore, according to the Tipler's definition, the naked singularity is gravitationally strong. The results for $8\pi(\phi_{,\lambda})^2[=\tau^{2}R_{\alpha\beta}T^{\alpha}T^{\beta}]$ versus $\lambda(\equiv-\ln\tau)$, plotted in the lower panel of Fig.~\ref{fig:stength}, show that along the geodesic $|\phi_{,\lambda}|$ is bounded. Then the integral $\int_{-\ln c}^{+\infty}(\phi_{,\lambda})^2\text{d}\lambda$ diverges. So by this means one also concludes that the singularity is strong.

It may be instructive to compare the results of the discrete self-similar critical collapse (Choptuik solution) with those of the continuous one (Roberts solution). In the Roberts solution, the spacetime at the center is not conformally flat~\cite{Guo_2018}. Moreover, in the $v<0$ region, the mass is negative. To avoid the negative mass problem, the spacetime of this region is replaced by a flat spacetime~\cite{Oshiro:1994hd,Brady:1993np}. The central singularity is weak~\cite{Nolan:1999tw}. It was commented that this singularity is actually a collapsed cone singularity rather than a naked singularity~\cite{Christodoulou_1994,Wald:1997wa}.

Since it was found that a naked singularity arises in Choptuik critical collapse, it has been argued that the initial data are not generic and so this model is not a serious conterexample to the Weak Comic Censorship Conjecture. Now it is observed here that, although critical collapse is a fine-tuning problem, the naked singularity forming in this process is gravitationally strong. The spacetime therefore cannot be extended beyond this singularity. So the physical nature of critical collapse may need to be examined further in light of this phenomenon.

\section*{Acknowledgments}
The authors are grateful to Xinliang An, Yun-Kau Lau, Junbin Li, Daoyan Wang and Xiaoning Wu for useful discussions. JQG thanks the Morningside Center of Mathematics, the Academy of Mathematics and Systems Science of the Chinese Academy of Sciences for hospitality. This work is supported by Shandong Province Natural Science Foundation under grant No.ZR2019MA068.


\end{document}